\documentclass{optica-article}

\journal{opticajournal} 

\articletype{Research Article}

\usepackage{lineno}
\usepackage{ulem}

\newcommand{\Br}{{\bf r}}
\newcommand{\Bk}{{\bf k}}

\begin{document}

\title{Deep learning and random light structuring ensure robust free-space communications}

\author{Xiaofei Li,\authormark{1,2}, Yu Wang,\authormark{3}, Xin Liu,\authormark{1}, Yuan Ma,\authormark{2}, Yangjian Cai,\authormark{1,*} Sergey A. Ponomarenko\authormark{2,4,*} and Xianlong Liu,\authormark{1,*}}

\address{\authormark{1}Shandong Provincial Engineering and Technical Center of Light Manipulation and Shandong Provincial Key Laboratory of Optics
and Photonic Devices, School of Physics and Electronics, Shandong Normal University, Jinan 250014, China\\
\authormark{2}Department of Electrical and
Computer Engineering, Dalhousie University, Halifax, Nova Scotia,
B3J 2X4, Canada\\
\authormark{3}School of Information Science and Engineering,Shandong Normal University, Jinan 250014, China\\
\authormark{4}Department of Physics and Atmospheric
Science, Dalhousie University, Halifax, Nova Scotia, B3H 4R2,
Canada}

\email{\authormark{*}yangjiancai@sdnu.edu.cn; serpo@dal.ca; xianlongliu@sdnu.edu.cn } 


\begin{abstract*} 
Having shown early promise, free-space optical communications (FSO) face formidable challenges in the age of information explosion. The ever-growing demand for greater channel communication capacity is one of the challenges. The inter-channel crosstalk, which severely degrades the quality of transmitted information, creates another roadblock in the way of efficient  FSO implementation. Here we advance theoretically and realize experimentally a potentially high-capacity FSO protocol that enables high-fidelity transfer of an image, or set of images through a complex environment. In our protocol, we complement random light structuring at the transmitter with a deep learning image classification platform at the receiver. Multiplexing novel, independent, mutually orthogonal degrees of freedom available to structured random light can potentially significantly boost the channel communication capacity of our protocol without introducing any deleterious crosstalk. Specifically, we show how one can multiplex the degrees of freedom associated with the source coherence radius and a spatial position of a beamlet within an array of structured random beams to greatly enhance the capacity of our communication link. The superb resilience of structured random light to environmental noise, as well as extreme efficiency of deep learning networks at classifying images guarantees high-fidelity image transfer within the framework of our protocol. 
\end{abstract*}

\section{Introduction}
In the age of information explosion, there has been growing demand for high capacity communication systems due to relentless growth of traffic through any available communication channel~\cite{YO23}. To achieve high-capacity communications within the framework of free-space optics, it is essential that all available degrees of freedom of a light field be explored and engaged. To date, a number of approaches to boost the transmission capacity of free-space optical communications have been proposed~\cite{IK18, ZK22, ZH23}, including a quadrature phase shift keying~\cite{BS22}, wavelength division multiplexing~\cite{DZ23}, space division multiplexing~\cite{CH22}, and polarization division multiplexing~\cite{AS23}. To further increase the channel capacity to overcome the existing bottle necks, the orbital angular momentum  division multiplexing has lately been proposed~\cite{AE20, JW22,JL23} . Apart from the capacity limitations, though, free-space optical communications  are hampered by the transmission quality degradation owning to the data crosstalk~\cite{HC22}. The latter is chiefly due to the challenge posed by a realistic transmission medium involving atmospheric turbulence and/or solid particles, such as aerosols, in the light path from a source (transmitter) to a receiver~\cite{JW22}.  Indeed, the refractive index of the atmosphere fluctuates due to the temperature and humidity variations, giving rise to turbulent effects. The atmospheric fluctuations distort the phase of a transmitted light beam, causing deleterious crosstalk among multiple independent degrees of freedom (DoF) of a free-space communication link~\cite{HC22}. As atmospheric turbulence seriously hinders further progress toward high-quality, high-capacity  optical communications through a realistic complex environment, numerous strategies have been proposed to remedy the channel crosstalk~\cite{ZX21,HS22,GM22,HZ23}.  In addition to multiple adaptive optics techniques, which can be subdivided into two main groups, pre-compensation and post-compensation~\cite{JW22}, there is a spatial polarization differential phase shift keying technology for vector light beams~\cite{ZZ21} and scattering-matrix-assisted retrieval protocol~\cite{LG19}. Unfortunately, virtually all these strategies require complicated, and often time-consuming,  data processing and they invariably fail to lower crosstalk to an acceptable level. Further, to compensate for light scattering obstacles in the transmission path, self-healing coherent light sources, such as the ones generating Bessel beams, have been employed to help self-reconstruct transmitted image structure past opaque obstructions~\cite{SL17,YS22}. 

At the same time, recent work has established extraordinary resistance of structured random beams to atmospheric turbulence and their outstanding self-healing ability upon encountering obstacles~\cite{Fei15,FW16,ZX20,ZX22}. In particular, the authors of~\cite{ZX22} have demonstrated that there exist  structured random beams maintaining their intensity profile structure in the turbulent atmosphere over a distance determined by the turbulence strength.  In contrast to fully spatially coherent fields~\cite{PSA03},  structured random fields possess a new degree of freedom, the normalized autocorrelation function of the fields at a pair of points across the source, known as the degree of coherence of a source~\cite{MW,PSA99}. This DoF has been explored to realize high-security optical data storage and retrieval~\cite{ArXiv}. Moreover, various aspects of optical field correlations at the source, for example, their spatial structure, transverse coherence radius, or classical entanglement~\cite{PSA21} provide access to numerous, untapped, mutually orthogonal DoFs that can be employed for high-capacity, high-fidelity optical communications through complex environments. 

In this work, we combine structured random light engineering at the source to encode an image into the novel DoFs of such light and deep learning framework at the receiver to propose theoretically and realize experimentally high-fidelity image transmission through a complex environment.  In particular, we employ statistically homogeneous, Laguerre-Gaussian correlated sources which produce optical fields with ring correlation structure at the source and ring-like far-field intensity profiles. We demonstrate that such ring-like patterns can be utilized to encode image information. We also show how the source coherence radius can be used as another independent DoF for information encoding. Further, we demonstrate experimentally that supreme resilience of structured random light to atmospheric fluctuations and their excellent ability to self-heal upon encountering obstacles augurs well for the fidelity of image transfer, at least, over short free-space communication links. We reveal the multiplexing capabilities of novel DoFs of random light. For instance, we show how the source coherence radius can be multiplexed with the space position of any beamlet within an array of partially coherent beams carrying image information to the receiver. These multiplexing capabilities demonstrate the potential for significant enhancement of information capacity within our protocol. 

Yet another novelty of our protocol is the use of deep learning network capabilities for pattern classification at the decoding stage. In recent years, deep learning has enjoyed immense success in computer science, making it possible to advance data-driven artificial intelligence technologies, such as computer vision~\cite{DL1}, speech recognition~\cite{DL2}, and decision making~\cite{DL3}. Recent applications of deep learning in photonics range from accurate prediction of resonance spectra~\cite{DL4} and inverse design of photonic devices~\cite{DL5} to high-resolution retrieval of orbital angular momentum states~\cite{DL6} and accurate phase prediction for anisotropic digital coding metasurfaces~\cite{DL7}. Although a deep convolutional neural network (CNN), which can be applied to classification tasks~\cite{DL8}, is the most popular, recently proposed residual networks (ResNets), which can scale up to thousands of layers, have demonstrated excellent promise for classification tasks with low training errors~\cite{KH16}. For these reasons, we employ ResNet 34 for image decoding at the receiver end. The fusion of random light structuring at the transmitter and deep learning image classification at the receiver 
renders our protocol a promising candidate to realize high-fidelity optical image transmission through a noisy link with a potential to attain high communication channel capacity via tapping into novel crosstalk-free DoFs available to random light sources. As an added bonus, our work will undoubtedly inform further research in the topical field of deep learning network applications to photonics.

\section{Image transmission employing source correlation structure}
\subsection{Encoding information into source correlations}
Consider a structured random light field propagating along the $z$ axis. In the space-frequency representation, we can describe the second-order correlations of the fields at a pair of points $\Br_1$ and $\Br_2$ in the transverse plane of the source in terms of a cross-spectral density of the source. We can express the cross-spectral density of any physically realizable statistical source as~\cite{Gori07}
\begin{equation}\label{W0}
	W_0(\Br_1,\Br_2)=\int d\Bk\, p(\Bk) H^{\ast}(\Br_1, \Bk) H(\Br_2, \Bk).
\end{equation}
Here, $H(\Br,\Bk)$ is an arbitrary kernel at a temporal frequency $\omega$ and $p(\Bk)$ is a nonnegative spectral distribution function in the reciprocal $k$-space; we will drop any explicit dependence on $\omega$ hereafter. 

Let us now focus on statistically homogeneous light sources for which $H(\Br,\Bk)=\sqrt{I(\Br)}e^{i\Bk\cdot\Br}$, where $I(\Br)$ is a source intensity profile~\cite{Gori07}. It then follows at once from Eq.~(\ref{W0}) that the degree of coherence $\mu_0(\Br_1-\Br_2)$ of such a source, defined as a normalized second-order correlation function $\mu_0(\Br_1,\Br_2)=W_0(\Br_1,\Br_2)/\sqrt{I(\Br_1) I(\Br_2)}$~\cite{MW,PSA03}, is simply a Fourier transform of $p(\Bk)$.  Further, the far-field intensity profile of a low-coherence source has the same functional form as $p(\Bk)$~\cite {MW}. Therefore, the spatial structure of source correlations, or equivalently, its spectral density distribution $p(\Bk)$ can serve as an independent DoF for information encoding. We choose a Laguerre-Gaussian (LG) correlated Schell-model source as  a representative example~\cite{Liang19}.  The spectral density of such a source reads
\begin{equation}\label{p}
	p(\Bk)\propto (\Bk^2 \sigma_c^2 /2)^{|l|}\left[ L_p^{|l|}(\Bk^2 \sigma_c^2 /2)\right]^2 e^{-\Bk^2 \sigma_c^2/2}.
\end{equation}
Here $\sigma _c$ is a coherence radius of the source, $L_p^{\left| l \right|}$ stands for an associated Laguerre polynomial of azimuthal $l$ and radial $p$ indices, respectively. We can then generate a multitude of LG-correlated sources by varying $p$ and $l$. Further, one can easily verify that a Fourier transform of the spectral density of the LG-correlated source yields yet another LG-like spatial pattern.  
\begin{figure}[h!]
\centering
\fbox{\includegraphics[width=13cm]{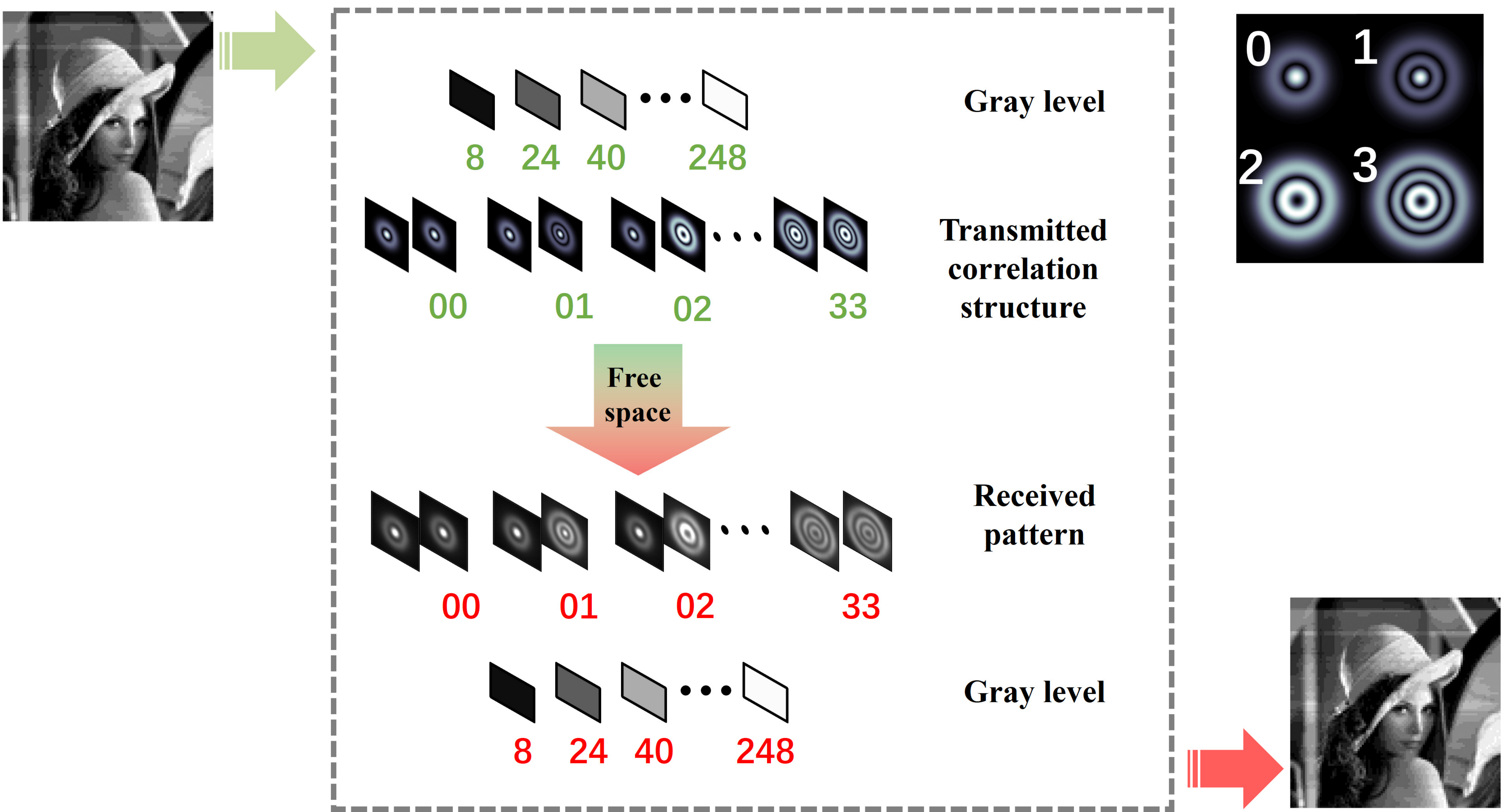}}
\caption{Schematics of our encoding/decoding protocol with LG correlated beams.}
\label{fig1}
\end{figure}

We are now in a position to describe our strategy to encode a desired image into an ensemble of light beams generated by an LG-correlated source. We sketch the schematics of our protocol in Fig.~\ref{fig1}. We exhibit 4 LG profile patterns, which we refer to as states hereafter, as an illustrative example of image encoding into LG spatial correlations of a source. The source produces an ensemble of LG correlated beams of spot size  $w=0.68$mm and coherence radius $\sigma_c=0.1$mm, propagating from the source (transmitter) to a receiver. The ensemble contains LG states corresponding to four index parameters: LG$_{10}$, LG$_{20}$, LG$_{12}$, and LG$_{22}$. The  transmitted information in our protocol is  a 16-grey-level  (8, 24, 40, 56, 72, 88, 104, 120, 136, 152, 168, 184, 200, 216, 232, and 248)  image of Lena which has $100\times100$ pixel resolution.  The grey level  of the image is quantified by 2-digit quaternary numbers (00, 01, 02, 03... 32, 33). Here, every quaternary number corresponds to an LG correlated beam structure (0 is LG$_{10}$, 1 is LG$_{20}$, 2 is LG$_{12}$, and 3 is LG$_{22}$). Therefore, we can encode the grey level of each pixel at the transmitter and store it in 2 LG correlated beam structures. Next, we transmit the encoded information from the source to the receiver in free space, or realistically through a random medium, with the aid of an ensemble of LG correlated beams. In the case of free-space propagation, the intensity patterns at the receiver, situated in the far-zone of the source,  are proportional to $p(\Bk)$ which is evaluated at $\Br$. Further, we record the intensity patterns of the received beam ensemble with a CCD camera and  decode them into a set  of quaternary numbers following  the encoding rules. Finally, we collect a set of pixels of variable grey level encapsulating the transmitted image.
\begin{figure}[htbp]
\centering
\fbox{\includegraphics[width=12.5cm]{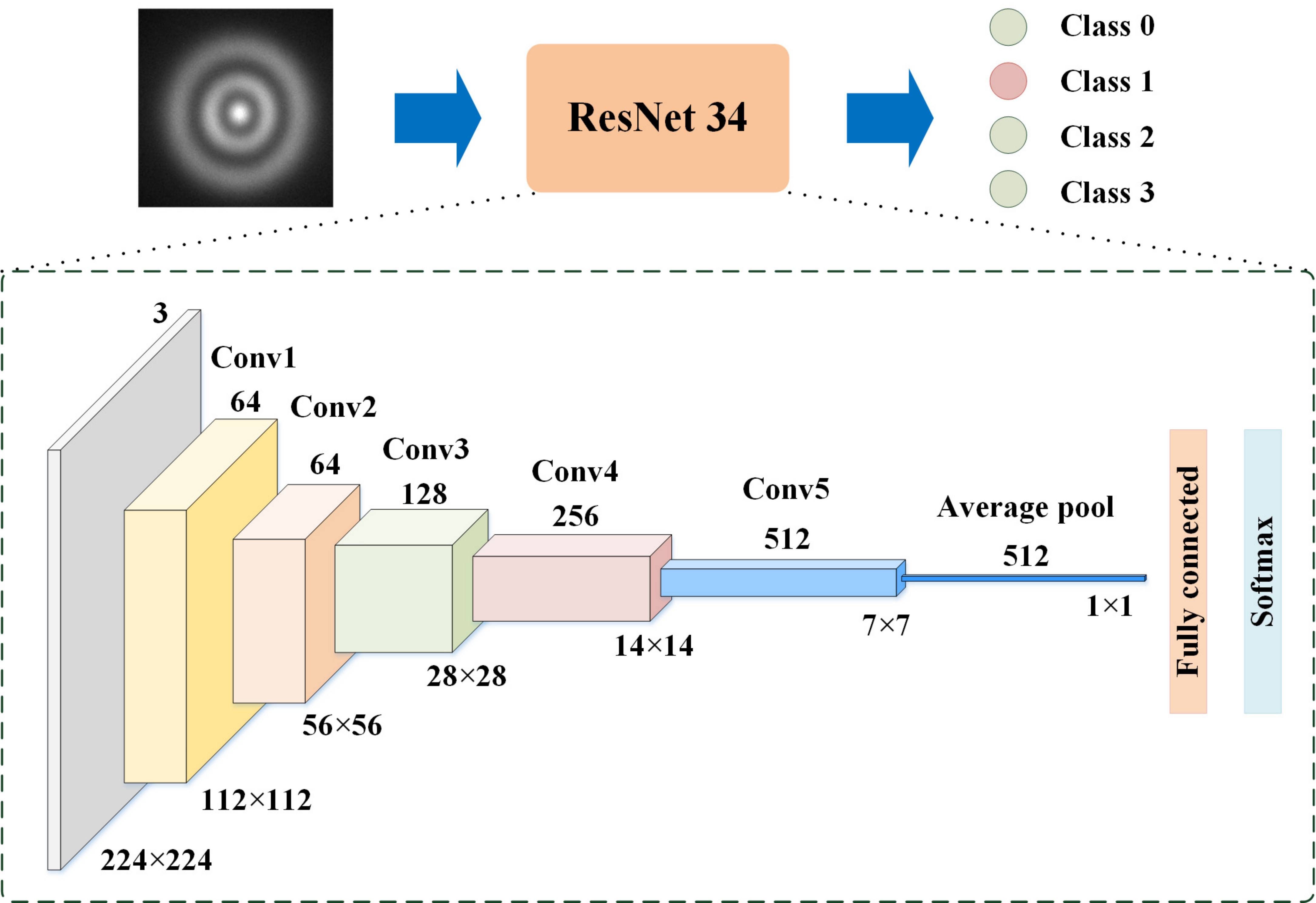}}
\caption{Decoding process and the architecture of ResNet 34 classifier.}
\label{fig2}
\end{figure}
\subsection{Decoding information with deep residual  learning} We employ a deep residual learning ResNet architecture with 34 layers, known as ResNet 34 network. The network is only interested in the image of intensity patterns and these patterns are directly accessible to the computers. First, we order sequentially all members of an LG-correlated beam ensemble recorded by the camera. In the following step, we have the network classify the received images and map them into quaternary numbers for subsequent decoding. Thus, each intensity pattern can be converted into a set of digital numbers by the classifier network  as shown in Fig.~\ref{fig2}. We display the basic framework of ResNet 34 classifier in the dashed square of Fig.~\ref{fig2} and we discuss the details of the overall structure of the network in Section 1 of the Supplement, see especially Tab.~S1 and Fig.~S1. We now outline how ResNet 34  works in a classifier mode. Inspired by the concept of transfer learning, we use the pre-trained parameters, which have been previously trained and tested on the ImageNet dataset\cite{OR15}, for initialization. We then fine-tune ResNet 34 with our images to suit a particular classification task. At the outset, we prepare  200 experimental images with labels for each category and combine them into a training dataset. Next we resize all the training images to $224\times224$ to be served as input to the network. To obtain a robust classifier model, we divide the whole training dataset randomly into three non-overlapping subsets in proportion $70\%$, $20\%$ and $10\%$, corresponding to the training, validation, and test sets, respectively. We utilize the training set to learn the characteristics of the data and develop a model. The validation set is used to adjust the parameters and hyper-parameters of the model during the training process to improve the model performance. The test set should not be used in the training phase and is used to evaluate the performance and robustness of the model on unseen data\cite{YC23}. Having taken these steps, we have established a robust ResNet 34 network ready to classify images. We then resize images to be tested and feed them into the established classification network, which realizes decoding in our protocol. Each tested image corresponds to the network output category, with the corresponding quaternary number specifying its grey level. Once the decimal grey value of each pixel is obtained, we can identify the whole image.

We employ a workstation with Ubuntu 18.04 OS, PyTorch platform, 32 GB RAM and NVIDIA Titan RTX for decoding. Due to the difference in the number of classes---while ImageNet has 1000 classes, we have only 4 classes---we only need to load the pre-trained parameters of the first 33 layers prior to training. The training is carried out by Adam optimizer\cite{KD14} with an initial learning rate of 0.0001 and a batch size of 32. The overall training phase requires 20 epochs. The training is carried out in an end-to-end fashion. 

\subsection{Experimental realization and transmission through realistic random medium}
\begin{figure}[htbp]
\centering
\fbox{\includegraphics[width=12cm]{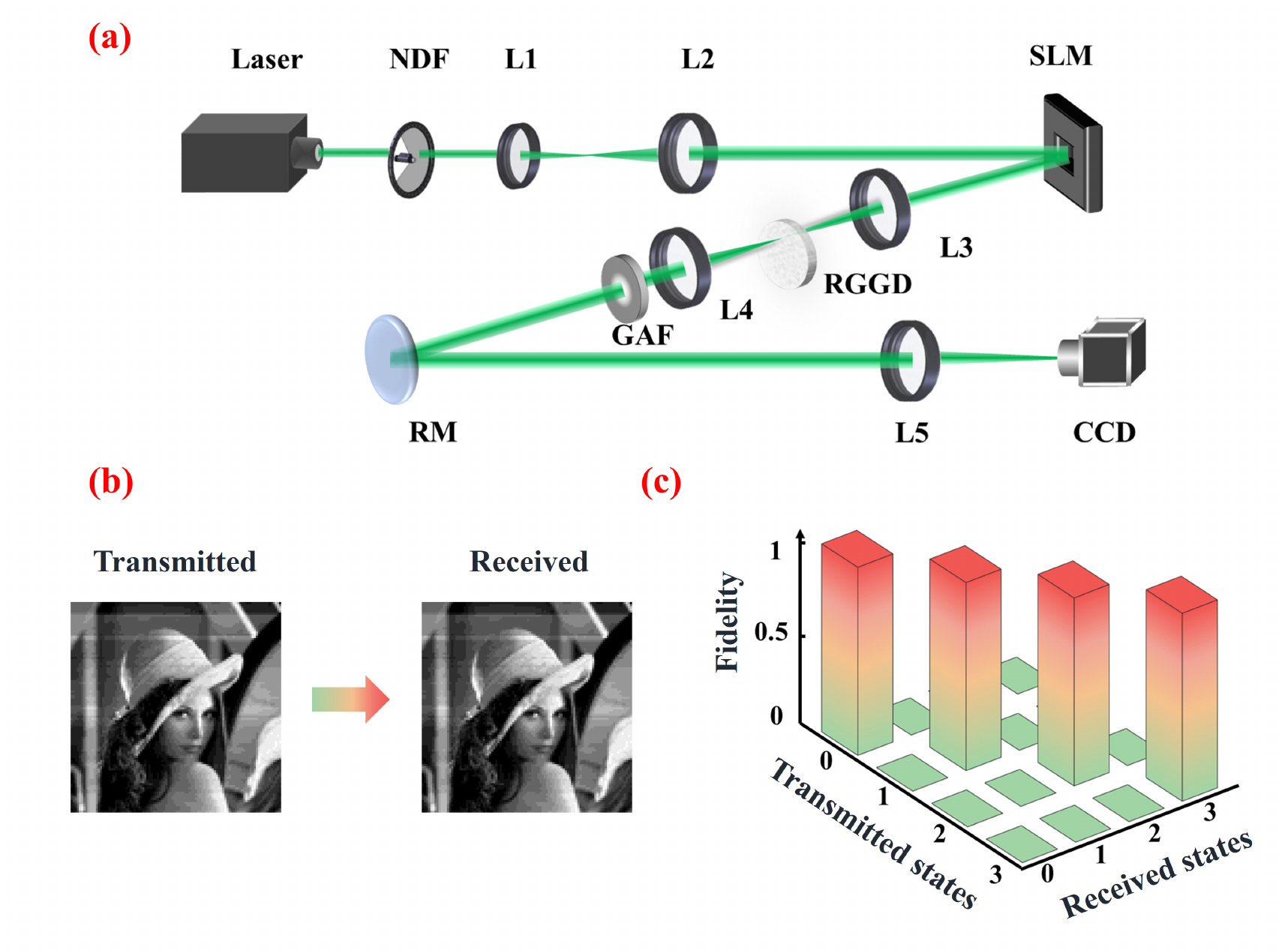}}
\caption{Experimental verification of our protocol. (a) Experimental setup. NDF: neutral density filter; L: thin lens; RGGD: rotating ground glass disk; GAF: Gaussian amplitude filter; SLM: spatial light modulator; RM: reflected mirror; and CCD: charge coupled device. (b) Qualitative juxtaposition of transmitted and received images. (c) Quantitative measure of the state detection fidelity as a conditional probability $P_{s_r \mid s_t }$ of finding a transmitted state $s_t$ in state $s_r$ , where  
$s_r  (s_t)$ is represented in the quaternary basis (0, 1, 2,  3). }

\label{fig3}
\end{figure}
We verify our protocol by carrying out proof-of-principle image transmission with an experimental setup sketched in Fig.~\ref{fig3}(a). First, we produce a fully coherent Gaussian beam with the help of a continuous-wave-diode-pumped laser with the carrier wavelength $\lambda_0=532$ nm and transmit the beam through a neutral density filter (NDF). After having been expanded and collimated by a beam expander (BE) comprised of two lenses of different focal length, L$_1$ and L$_2$,  the beam is reflected by a spatial light modulator (SLM), which displays a hologram dataset of LG patterns encapsulating the encoded information. Thereby generated LG beams are focused by a lens L$_3$ ($f=250$mm) and projected onto a rotating ground glass disk (RGGD), which we place in the front focal plane of a collimating lens L$_4$. Next, an LG-correlated beam emerges past L$_4$ ($f=150$mm) and a Gaussian amplitude Filter (GAF) and we encode the transmitted information into its spatial correlation structure. Having  propagated a $1.92$m stretch of free space, the  beam is focused onto a charge coupled device (CCD) camera by a lens L$_5$ ($f=150$mm). The camera serves as a receiver recording the intensity patterns of an LG-correlated  beam ensemble.  

We contrast the transmitted  and received images of Lena in {Fig.~\ref{fig3}(b). We can readily infer from the figure that there is excellent qualitative resemblance between  the transmitted and received images attesting to the viability of our protocol.  We then quantify the fidelity of each state in terms of a conditional probability $P_{s_r \mid s_t }$  of finding a transmitted state $s_t$ in state $s_r$~\cite{ZZ21} .  We summarize the results in Fig.~\ref{fig3}(c).} The multitude of $100\times100\times2=20000$ received states are recognized by the ResNet 34 network, and the attained  fidelity of the image transfer, evaluated as a fraction of correctly received states versus the total number of transmitted states, exceeds $99.99\%$.
\begin{figure}[htbp]
\centering
\fbox{\includegraphics[width=11cm]{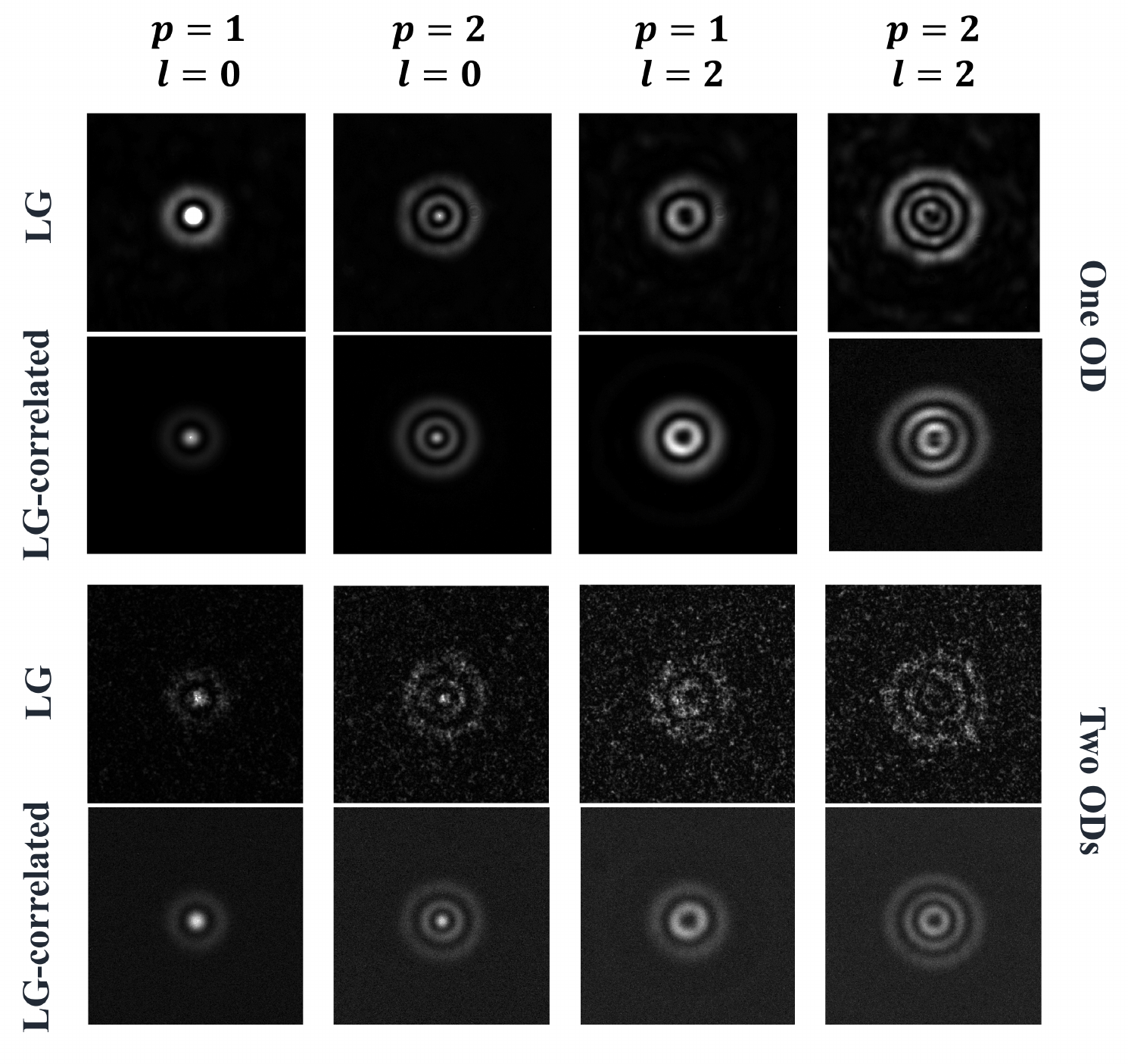}}
\caption{Recorded intensity profiles of LG-correlated versus fully coherent LG beams transmitted through either a single or two optical diffusers.}
\label{fig4}
\end{figure}

We now elucidate the role of noise due to either atmospheric turbulence or opaque obstacle scattering on image transmission quality within the framework of our protocol. There has been a growing body of evidence, see, for instance, \cite{Fei15} for a review that decreasing spatial coherence of a light source enhances the resilience of the beams  generated by such a partially coherent source to  random perturbations in a complex environment. In particular, it has been recently demonstrated~\cite{ZX22} that there exists a class of structured random beams, the so-call dark and antidark beams on incoherent background~\cite{DAD07} that maintain their intensity profile structure over certain propagation distances in the turbulent atmosphere,  regardless of the turbulence strength. Instructively, the intensity profile of any such structurally stable beam has a ring structure, similar to that of the LG-correlated beams which we employ in our protocol. Therefore, we anticipate the latter to be fairly robust against atmospheric turbulence as well. To verify this conjecture, we record the intensity patterns of LG-correlated beams propagating through a random medium mimicking atmospheric turbulence and compare them with those of fully coherent LG beams propagating through the same medium.  We use optical diffusers (OD) to simulate turbulence, see the Supplement for further information. We can infer from Fig.~\ref{fig4}, that  fully coherent LG beams  are more distorted than their partially coherent  cousins. This observation implies that  image transmission through random medium and recovery of fully coherent LG beams are much more complicated and time consuming as well as far less accurate than those with LG correlated  beams. Moreover, the situation exacerbates as the turbulence strength increases. We can model stronger turbulence with two ODs as opposed to weaker turbulence which can be modelled with just a single diffuser. By comparing the corresponding panels of Fig. 4, we observe that the advantage of reducing source coherence of the beam augments for stronger turbulence: We clearly distinguish a ring structure of the LG-correlated beam in stronger turbulence, while the intensity profile of a fully coherent LG beam swiftly turns into essentially a random speckle pattern. 

To test our protocol in the more adverse situation,  we repeat all protocol steps, except we include two ODs to model stronger turbulence as a propagation milieu. We transmit  a 16-grey-level Lena image of $80\times80$ pixel resolution.  We present an experimental setup schematics in Fig.~S2 of the Supplement.  Our results show good tolerance of the protocol to medium turbulence and yield the fidelity of image retrieval of over $99.99\%$, which can be inferred upon comparison of the transmitted and received states in Fig.~2(b). 

By the same token,  random beams of sufficiently low coherence are known to self-heal upon encountering  discrete obstacles such as suspended particles in free space~\cite{FW16,liu17}.  In Fig.~S3 of the Supplement, we provide experimental evidence of LG-correlated beam self-healing capabilities by comparing its evolution with that of a fully coherent LG beam after the cross-section of either beam has been partly blocked by an opaque obstacle. We then repeat  our protocol for a 16-grey-level Lena image of $86\times86$ pixel resolution and are able to accurately transmit the image through a medium containing obstructing particles to the fidelity of over $99.99\%$, see also Fig.~S6.  We remark that the extremely high fidelity of image transmission in free-space and turbulent or colloidal particle medium is achieved thanks to the superb job that a deep learning network, such as ResNet 34, does to faithfully decode images. 
\section{Image transmission utilizing source coherence radius}
\begin{figure}[htbp]
\centering
\fbox{\includegraphics[width=12cm]{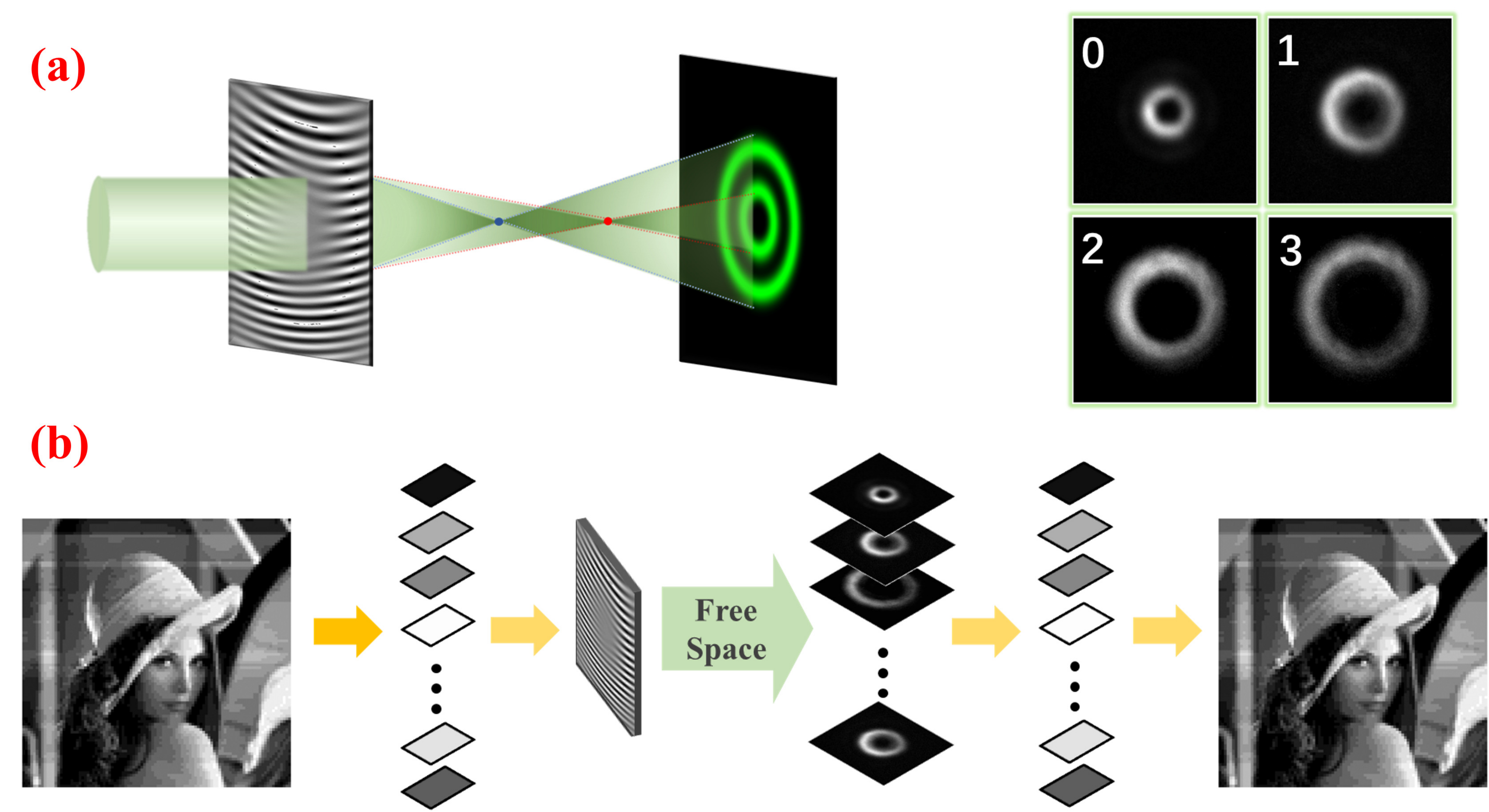}}
\caption{Schematics of the image transfer protocol employing the source coherence radius for information encoding. (a) Illustrating control of the transverse coherence radius of the source with a focusing hologram. (b) The image transmission protocol involving the spatial coherence radius DoF. }
\label{fig5}
\end{figure}

We now illustrate an alternative strategy of encoding image information into the transverse coherence radius of a light source which can be viewed as another independent degree of freedom for free-space optical communications.  To be consistent, we employ an LG-correlated source as well, so that we can encode the Lena image into coherence radii of LG ring patterns. To this end, we select 4 distinct states of an LG-correlated beam with the ring structure corresponding to LG$_{04}$ of the same spot size $w=0.68$mm and the coherence radii at the source $\sigma_c=0.02$mm, $0.025$mm, $0.03$mm, $0.05$mm  as an example. Notice that all four states have low coherence in the sense that $\sigma_c \ll w_0$.  In the protocol described in Section 2,  we impart a phase $\Phi _{LG}$ with a phase-only hologram to generate an LG-correlated source. The source coherence radius depends on the spot size of the coherent beam incident on the RGGD. The spot size is, in turn, controlled by adjusting the distance between L$_3$ and RGGD. In the alternative strategy, we should be able to control the coherence radius of the source without affecting the other parameters of the experimental setup. With this purpose, we employ a digital focusing hologram~\cite{CRb17} to tune the coherent beam spot size digitally without affecting any distances between optical elements. It follows that for a fixed distance between SLM and RGGD, the spot size is controlled by the focal length $f$ of the hologram as we show in Fig~\ref{fig5}(a). As a result, the coherence radius of the LG-correlated source can be modified. Thus, the source coherence can be modulated by refreshing holograms on SLM. The transfer function of the digital hologram is a combination of that of a focusing element, $t\left( {x,y} \right) = \exp [i\frac{k_0}{{2f}}({x^2} + {y^2})]$, where $k_0=2\pi/\lambda_0$ and the transfer function of a linear phase grating, ${t_{\rm{g}}}\left( {x,y} \right) = \exp [i2\pi \left( {ux + vy} \right)]$, where $u$ and $v$ are inverse grating periods along the $x$- and $y$-directions, respectively. Therefore, the hologram for image encoding into the source coherence radius modifies the phase of an input beam to
 \begin{equation}
\begin{split}
\label{Phi}
	\Phi  = \bmod \left[ {{\Phi _{LG}} + \frac{k_0}{{2f}}({x^2} + {y^2}) + 2\pi \left( {ux + vy} \right), 2\pi } \right].
\end{split}
\end{equation}

 In Fig.~\ref{fig5}(b), we  illustrate the principle of our alternative image transmission protocol. Specifically, we encode a 16 grey-level Lena image of $90\times90$ pixel resolution into 4 states, corresponding to the LG-correlated sources of different coherence radii: 0 is $\sigma_c=0.05$mm, 1 is $\sigma_c=0.03$mm, 2 is $\sigma_c=0.025$mm, 3 is $\sigma_c=0.02$mm. The grey level value of each pixel is thus encoded into the coherence radius of an LG-correlated beam generated by the corresponding source. Following this procedure, we generate an ensemble of LG-correlated beams propagating in free space from the source to the receiver. At the receiver, we capture and decode  $90\times90\times2=16200$ intensity profile images with the aid of a CCD camera and ResNet 34 network. Finally, we collect the grey level values of all image pixels and recover the image of Lena. As is evidenced by Fig.~\ref{fig5}(b), the received image bears excellent resemblance to the transmitted one, attesting to the viability of our protocol. 
 
The two protocols that we have described in Sections 2 and 3  involve two independent DoFs for free-space optical communications: the spatial structure of the source degree of coherence and the coherence radius of the source. Each of these DoFs can be multiplexed with other optical field characteristics such as spatial, polarization, and spectral (wavelength) DoFs to boost communication capacity of the protocol. 
\begin{figure}[h]
\centering
\fbox{\includegraphics[width=12cm]{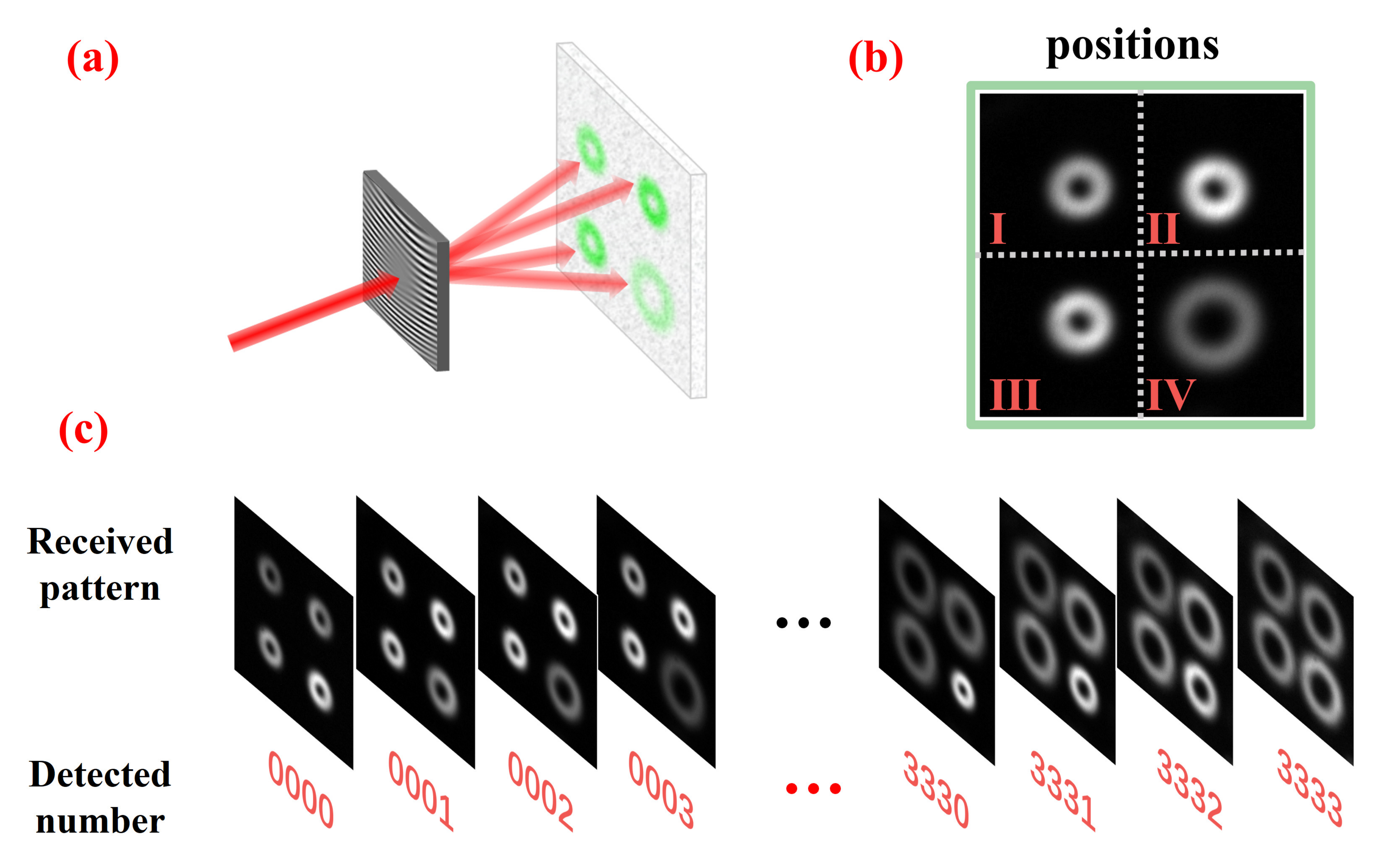}}
\caption{Schematics of the image transfer protocol employing an array of LG-correlated beams.  (a) Schematic representation of the array generation. (b) The position of each intensity pattern in a $2\times2$ array. (c) The encoding/decoding sequence of 256 grey levels and 256 array states.}
\label{fig6}
\end{figure}
\begin{figure}[h!]
\centering
\fbox{\includegraphics[width=13cm]{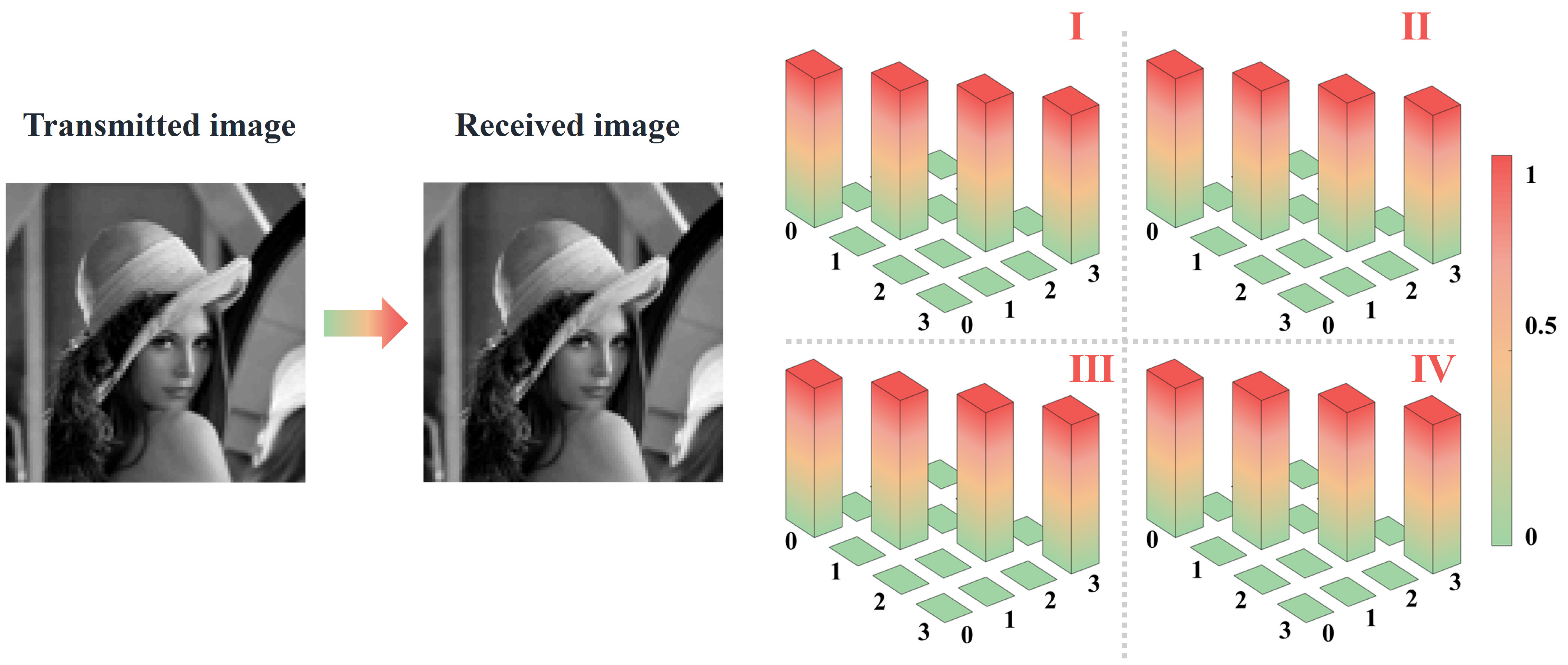}}
\caption{ Qualitative (visual) and quantitative (fidelity) comparison between transmitted and received images. The color bar shows normalized fidelity with unity corresponding to $100\%$.}
\label{fig7}
\end{figure}

We illustrate multiplexing the coherence radius and spatial location of a beamlet within a source array to illustrate but one among multiple perspectives. We encode information at the transmitter by generating an array of LG correlated beams with the RGGD. Conceptually, the idea of our multiplexing method stems from the spatial coherence radius modulation with the phase hologram realizing the transformation of Eq.~\ref{Phi}. By adjusting the periods of the linear grating ${t_{\rm{g}}}\left( {x,y} \right)$, we can steer each beamlet of the array to a distinct spatial location as is visualized in Fig.~\ref{fig6}(a). Thus, we are able to control the spatial coherence radius of any beamlet within the array.  Inspired by~\cite{CR17}, we can then jointly control the coherence radius and spatial location of the beamlet with the aid of a single hologram that performs a cumulative phase $\Phi_{\rm{M}}$ transformation viz., $\Phi_M =\sum_{j=1}^M \Phi_j$, where $\Phi_j$ is given by Eq.~(\ref{Phi}) with $f_j$, $u_j$ and $v_j$ corresponding to the $j$th state.  By introducing the space dimension into the encoding protocol, we greatly increase the number of available encoded states. Indeed, in the protocol sketched in  Fig.~\ref{fig6}, we engage 4 coherence state channels which correspond to four sources of variable coherence radius according to the rule: 0 is $\sigma_c=0.05$mm, 1 is $\sigma_c=0.03$mm, 2 is $\sigma_c=0.025$mm, 3 is $\sigma_c=0.02$mm. We multiplex each coherence state channel with 4 position channels (\uppercase\expandafter{\romannumeral1}, \uppercase\expandafter{\romannumeral2}, \uppercase\expandafter{\romannumeral3}, \uppercase\expandafter{\romannumeral4}), yielding $4^4=256$  states (0000, 0001, 0002 ... 3332, 3333). The transmitted image is a large-scale, grey Lena image of $100\times100$ pixel resolution corresponding to 256 grey levels (0-255). In this experiment, the grey level value of each pixel is represented by a single array of LG-correlated beams. After having propagated over a stretch of free space, the array intensity patterns are recorded by the CCD camera and mapped into the grey value diagram by ResNet 34. Finally, we compare the decoded and encoded grey-level information to ascertain the transmission fidelity to be $99.99\%$, see Fig.~\ref{fig7}.

\section{Conclusion}
We have theoretically proposed and experimentally implemented a free-space optical communication protocol enabling high-fidelity transfer of a desired image or a set of images through, in general, a noisy communication channel. The key novelty of the proposed protocol is the fusion of random source structuring at the transmitter end and the employment of a deep learning network at the receiver end. The structuring of random source makes it possible to tap into multiple hitherto unexplored DoFs for image encoding, including the two that have been explicitly demonstrated to yield high-fidelity information transfer: the coherence structure of the source and the source coherence radius. In addition, using structured random light ensures resilience of the transmitted image to channel noise such as medium turbulence or the presence of solid obstacles in the light propagation path. The high-fidelity of our protocol is achieved due to two factors. First, the structured random light is robust against random perturbations in the medium. Second and most important, the deep learning network, which we employ for image decoding, enables extremely accurate image recognition in a time efficient manner. We note in passing that a recently proposed alternative protocol~\cite{ArXiv}, which requires detailed knowledge of two-point field correlations at  the source, rather than just their generic structure or source coherence radius, is much more time consuming. 

We have demonstrated the capability of our protocol for multiple DoF multiplexing to enhance communication capacity. In particular, we have shown how to multiplex the source coherence radius DoF with that of a spatial position of a beamlet within a random beam array carrying the image information. At the moment, our protocol is limited to relatively short propagation links in a noisy environment. We conjecture that the incorporation of dark/antidark beams on incoherent background, which are known to maintain the ring structure of their intensity profiles over tens of kilometers in weak, and hundreds of meters in strong atmospheric turbulence~\cite{ZX21}, into our protocol can significantly extend the operational range of the latter without compromising the fidelity of transmitted images.

\begin{backmatter}
\bmsection{Funding}
National Key Research and Development Program of China (2022YFA1404800, 2019YFA0705000), National Natural Science Foundation of China (12192254, 92250304), China Scholarship Council (202108370219) and Natural Sciences and Engineering Research Council of Canada (RGPIN-2018-05497).

\bmsection{Disclosures}
\noindent  The authors declare no conflicts of interest.

\bmsection{Data Availability Statement}
The data that support the findings of this study are available from the corresponding authors upon reasonable request.

\bmsection{Supplemental document}
See Supplement for supporting content. 

\end{backmatter}







\end{document}


\title{Deep learning and random light structuring ensure robust free-space communications: supplemental document}
\maketitle
\section{The structure of ResNet 34}
\begin{table}[htbp]
\centering
\caption{\bf ResNet 34 architecture}
{\begin{tabular}{@{}ccc@{}} 
\toprule
Layer Name & Output Size & 34-Layer \\ 
\hline
Conv1  & $112\times112$ & $7\times7$, 64, stride 2  \\
Conv2  & $56\times56$   &  \makecell{$3\times3$ max pool, stride 2 \\ 
$\begin{bmatrix} 3\times3, & 64 \\ 3\times3, & 64 \end{bmatrix}\times3$}  \\
Conv3  & $28\times28$   & $\begin{bmatrix} 3\times3, & 128 \\ 3\times3, & 128 \end{bmatrix}\times4$   \\
Conv4  & $14\times14$   & $\begin{bmatrix} 3\times3, & 256 \\ 3\times3, & 256 \end{bmatrix}\times6$   \\
Conv5  & $7\times7$     & $\begin{bmatrix} 3\times3, & 512 \\ 3\times3, & 512 \end{bmatrix}\times3$   \\
       & $1\times1$     & average pool, fc, softmax   \\ 
\hline
\end{tabular}}
\label{tab1}
\end{table}
\begin{figure}[h]
\centering
\fbox{\includegraphics[width=3.5cm]{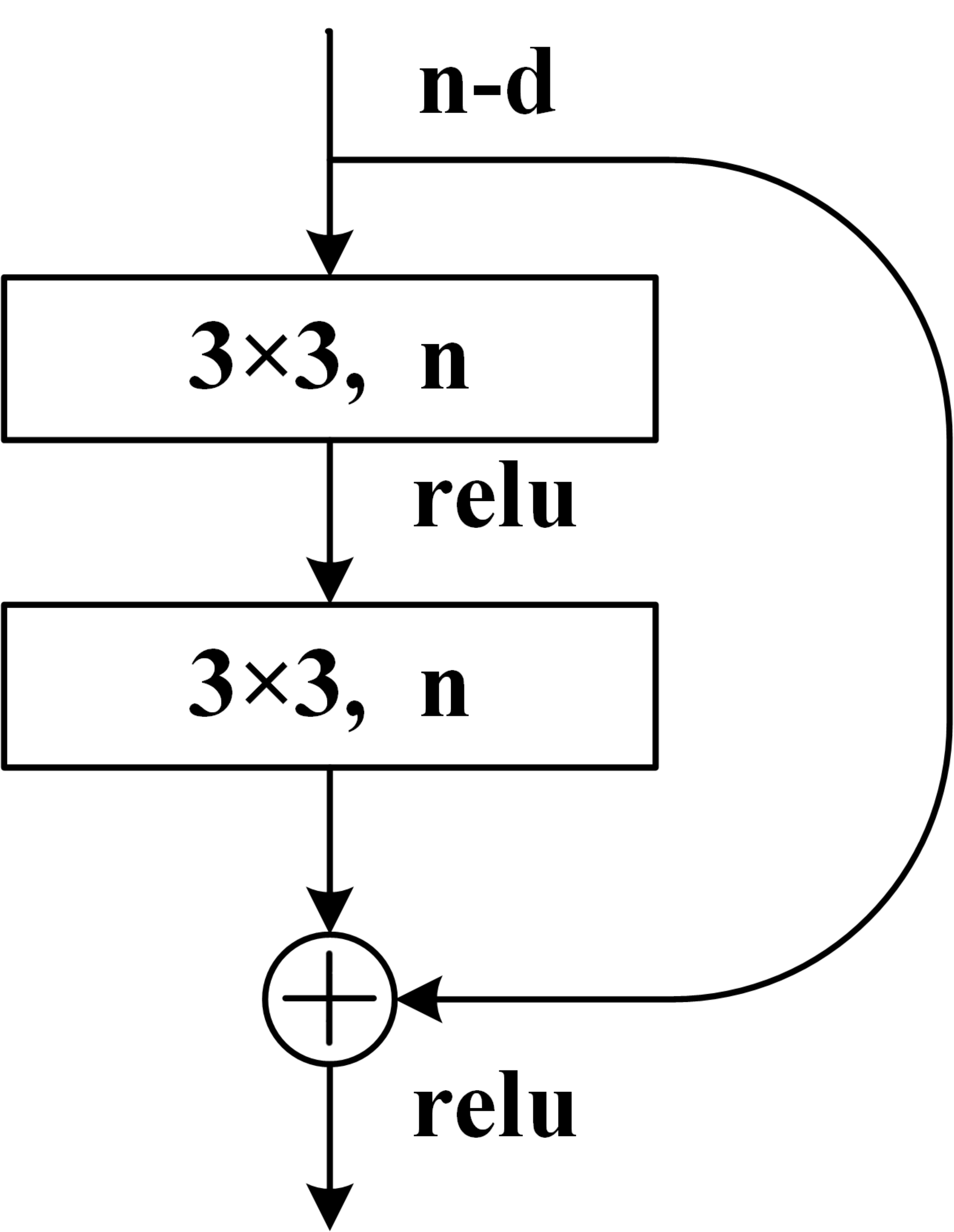}}
\caption{Residual building block diagram of ResNet 34 with $n$ representing a number of convolution kernels.}
\label{figs1}
\end{figure}
ResNet 34 \cite{he-16} is a specific implementation of a deep residual network; we list the ResNet34 parameters in Tab.~\ref{tab1}. ResNet 34  comprises 33 convolutional layers and one average pool layer,  followed by a fully connected layer. The overall architecture of ResNet 34 can be subdivided into five modules from Conv1 to Conv5. Conv1 is a $7\times7$ convolutional layer with a stride of $2$, followed by a $3\times3$ max pooling layer with a stride of $2$. The next four modules, Conv2 to Conv5, mainly consist of a series of residual building blocks which form the basis of ResNet 34 network. The residual building block is composed of the following convolutional layers: batch normalization, rectified linear unit activation function (ReLU) and a shortcut, as shown in Fig.~\ref{figs1}.
The residual block is designed to effectively alleviate the problem of vanishing and exploding gradients caused by the depth increase of the neural network. It can improve the classification efficiency, namely, the decoding fidelity of our FSO system. The number of residual building blocks embedded into Conv2, Conv3, Conv4 and Conv5 is 3, 4, 6 and 3, respectively. The convolution kernel size is fixed at $3\times3$. The corresponding numbers of convolution kernels are then 64, 128, 256 and 512. Additionally, after each stage, downsampling is performed to reduce the size of feature maps and achieved by a $1\times1$ convolutional layer with a stride of $2$. The final module of ResNet 34 is a global average pooling layer which transforms the feature map into a vector. A fully connected layer with a softmax activation function is then used to obtain the final classification results.

\section{Experimental setup for image transfer through random medium}
\begin{figure}[htbp]
\centering
\fbox{\includegraphics[width=11cm]{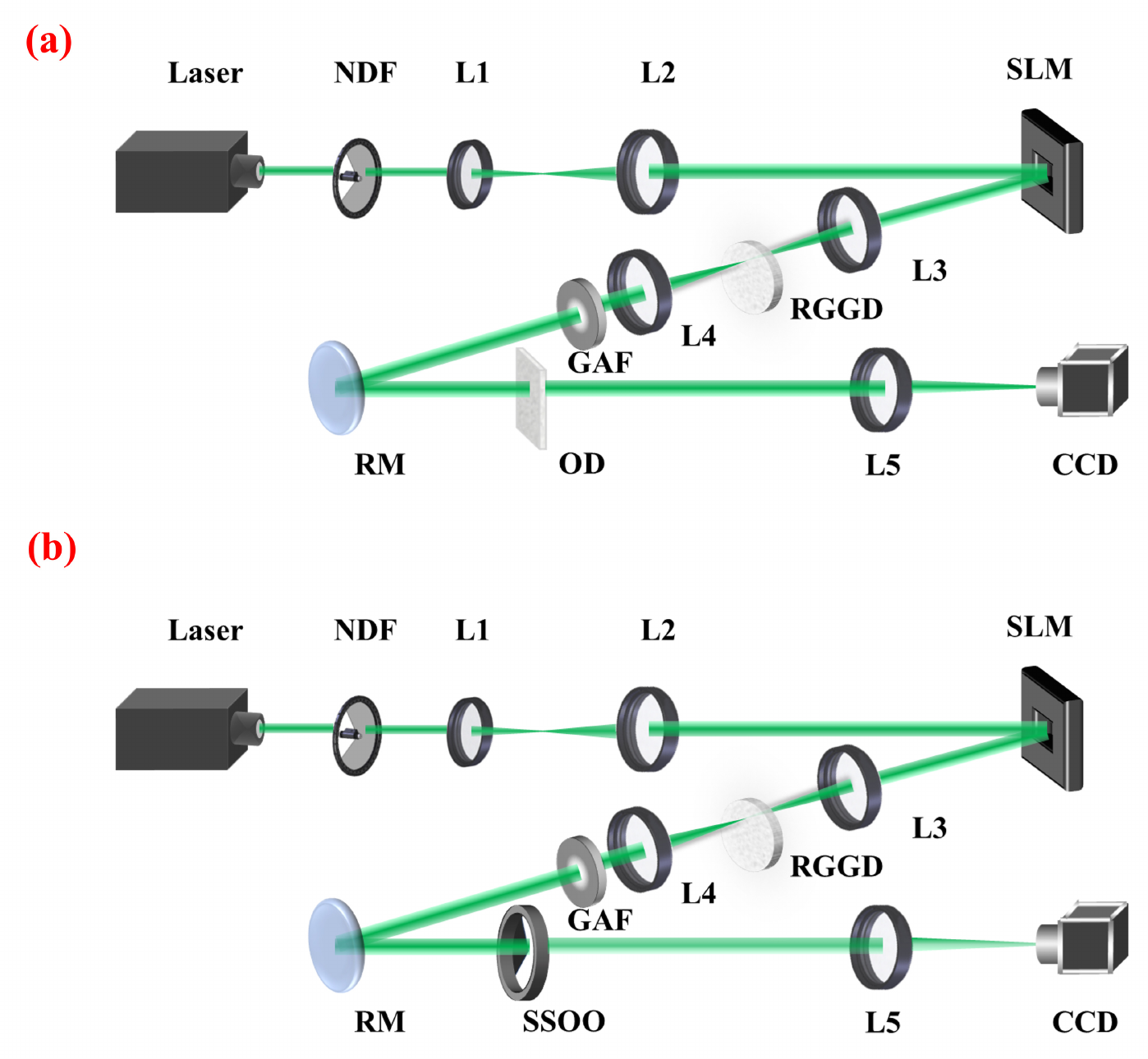}}
\caption{Experimental setup of grey image (a) transmitting in atmospheric turbulence and (b) suffering obstructing particles. OD: optical diffuser; SSOO: sector-shaped opaque obstacle, others are same with Fig.~3.}
\label{figs2}
\end{figure}
The setup for image transmission through a realistic random medium is similar to that in Fig.3; only the optical elements emulating the complex environment are added. In the image transmission through turbulence experiment---see Fig.~\ref{figs2}a---we apply an optical diffuser (OD) with a diffuse Gaussian-like distribution to emulate atmospheric turbulence (\url{https://www.lbtek.com/product/92/product_id/1489.html#DW120-1500}). A OD is employed to simulate weaker turbulence, while two ODs are placed in the beam path to mimic stronger turbulence. We place a sector-shaped opaque obstacle (SSOO) with an angle of 120\degree  in the light path to model obstructing particles in a separate experiment shown in Fig~\ref{figs2}b. 
\section{Resilience of LG-correlated beams to obstacles}
\begin{figure}[h]
\centering
\fbox{\includegraphics[width=9cm]{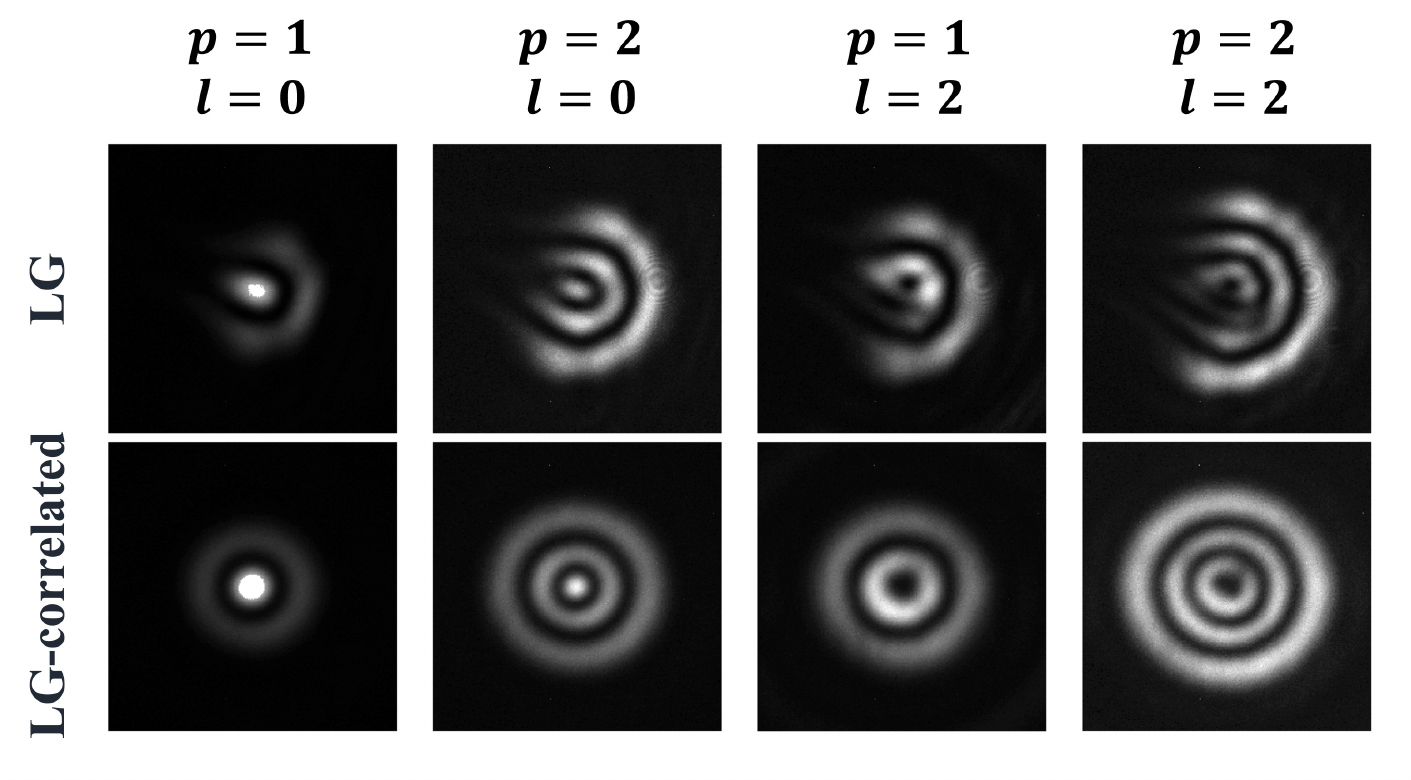}}
\caption{Recorded intensity profiles of LG-correlated versus fully coherent LG beams obstructed by a sector-shape obstacle situated in the beam path.}
\label{figs3}
\end{figure}

We can see in Fig.~\ref{figs3} that the structure of fully coherent LG beams at the receiver is heavily affected by the presence of solid particles in the medium, while the corresponding patterns of LG-correlated beams are hardly affected by the obstructions at all due to their outstanding self-healing capabilities. Our results are consistent with the previous work on the subject~\cite{liu-17,li-20}. 

\section{LG-correlated beam intensity versus source coherence radius}
In our protocol utilizing the source coherence radius for image encoding, the 4 quaternary states for image encoding correspond to $\sigma_c=0.02$mm, $0.025$mm, $0.03$mm, $0.05$mm. These states are realized with the aid of digital focusing holograms of four focal lengths: $f = 50, 500, 700, $ and $1200$mm.
\begin{figure}[h]
\centering
\fbox{\includegraphics[width=9cm]{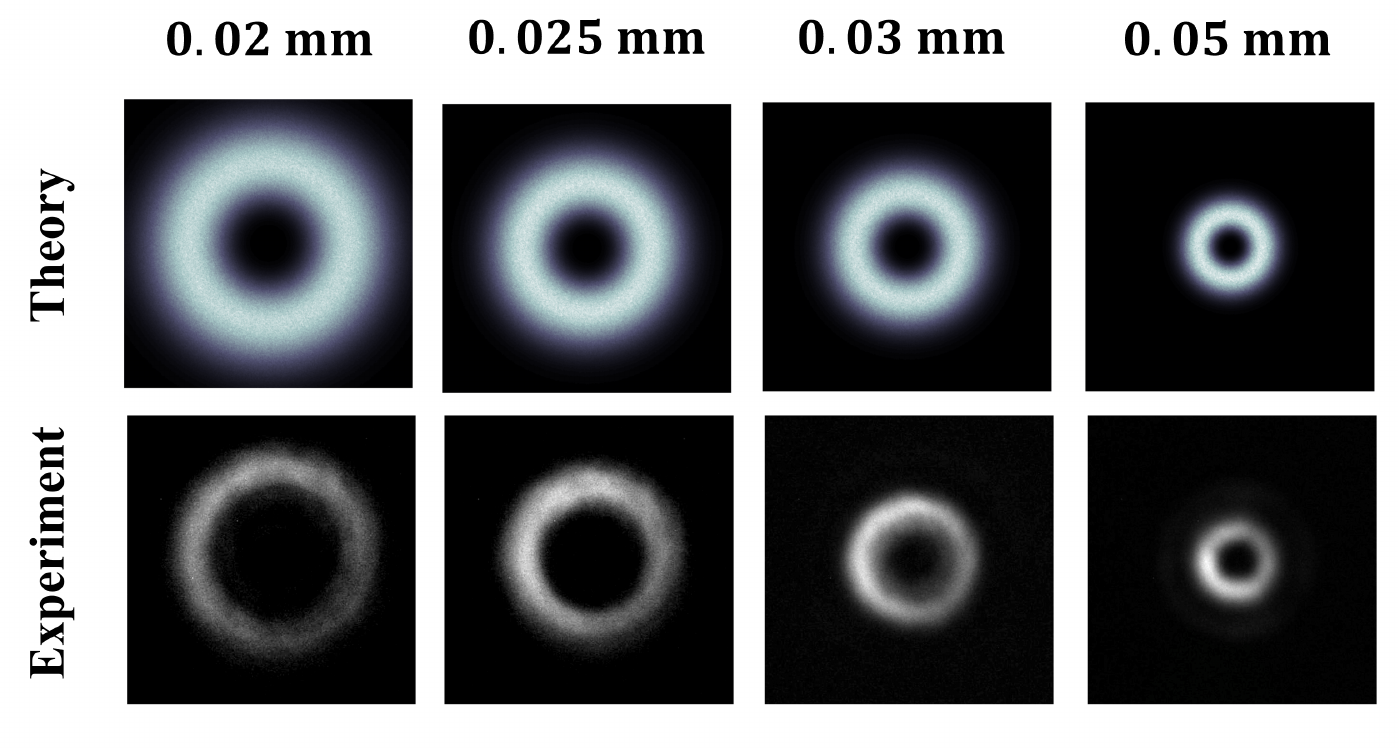}}
\caption{Normalized intensity of an LG-correlated beam (LG$_{04}$) at the receiver generated by a source of variable coherence radius. }
\label{figs4}
\end{figure}
\section{Transmitted image fidelity evaluation  }
\begin{figure}[h!]
\centering
\fbox{\includegraphics[width=11cm]{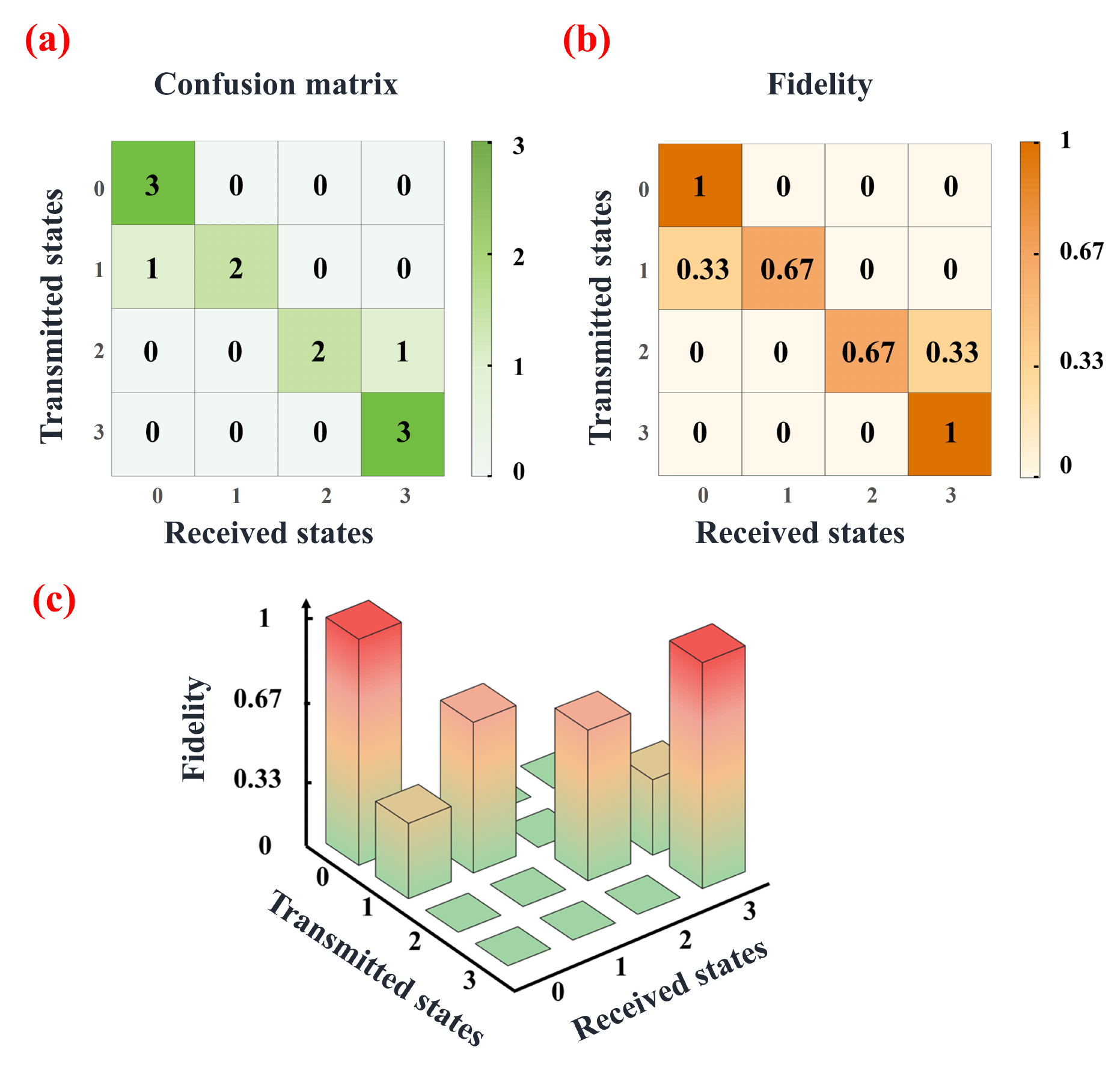}}
\caption{Confusion matrix and transmission fidelity of an illustrative example.}
\label{figs5}
\end{figure}
\begin{figure}[h!]
\centering
\fbox{\includegraphics[width=11cm]{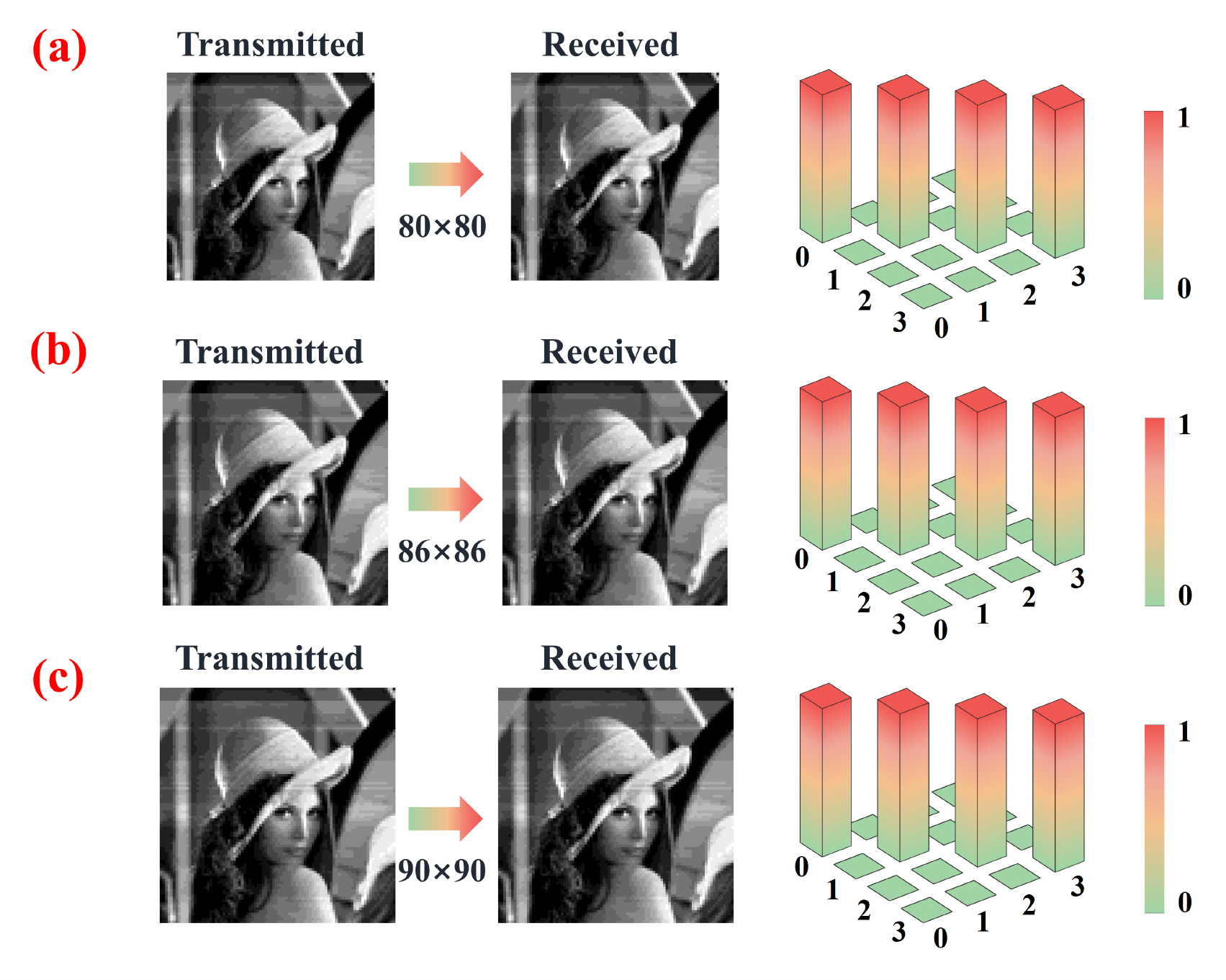}}
\caption{Qualitative (visual) and quantitative (fidelity) comparison between transmitted
and received images. (a), \& (b) correspond to the protocol utilizing the source correlation structure, while (c) corresponds to that using the source coherence length for image encoding. The color bar shows normalized fidelity with unity corresponding
to 100$\%$. }
\label{figs6}
\end{figure}

We assess the performance of our protocol by evaluating the conditional probability $P_{s_r \mid s_t }$ of a measuring state $s_t$ given $s_r$ and the total fidelity of each transmission. 
Here, we give a simple example to illustrate the evaluation process. We stress that this is just an illustrative example which has nothing to do with the actual transmitted data.

Suppose a transmitted data string is composed of 12 quaternary numbers, "000111222333", say.  Further, assume that the received string is "0001\textcolor{red}{0}122\textcolor{red}{3}333". First, we notice that each  transmitted quaternary state is composed of 3 digits, such that each transmitted state $s_t$ makes the fraction $P_{0 }=P_{1 }=P_{2 }=P_{3 }=3/12=0.25$ of the total 12 digits making up the information string. Further, on comparing the transmitted and received data strings, we can infer that 2 states are incorrectly received (marked in red ink):  the 5th digit in the received string is 0 instead of 1, while 9th is 3 in lieu of 2.  We can then compose a confusion matrix $N$ based on these observations. The  row and column elements of the confusion matrix are separately determined by the transmitted and received states. Starting with the first row, all 3 transmitted "0" are correctly received as "0", implying that  $N_{00}=3$, 
and $N_{01}=N_{02}=N_{03}=0$. Next, consider the second row. We notice that  3 transmitted states "1" are received as 2 states "1" (correct) and 1 state "0" (incorrect). Thus,  $N_{10}=1$, $N_{11}=2$, and $N_{12}=N_{13}=0$. Following the same procedure, we can fill up the other two rows of the ``green" matrix . We have $N_{22}=2$, $N_{23}=1$ and $N_{33}=3$, with the other  matrix elements being zeroes. We exhibit the resulting matrix in Fig.~\ref{figs5}a. We can then easily evaluate the conditional probabilities of of finding transmitted states in given received states as follows: $P_{0 \mid 0}=P_{3 \mid 3}=3/3=1$, $P_{1 \mid 1}=P_{2 \mid 2}=2/3=0.67$, $P_{0 \mid 1}=P_{3 \mid 2}=1/3=0.33$, with the others being 0. We display these results in the ``brown" matrix of Fig.~\ref{figs5}b, as well as show them with the bar chart in Fig.~\ref{figs5}c.  Finally, we can evaluate the total fidelity of data transmission as  $\sum\limits_{{s_r} = {s_t}} {P\left( {{s_r}\left| {{s_t}} \right.} \right)P\left( {{s_t}} \right)} =1\times0.25+\frac {2}{3}\times0.25+\frac {2}{3}\times0.25+1\times0.25=83.3\%$.

At the same time, the total fidelity of image transmission is determined by the fraction of correctly received states versus the total number of transmitted states as $10/12=83.3\%$, which is in complete agreement with our calculation based on state detection probabilities.
 
\section{Qualitative comparison of transmitted and received images}
We juxtapose the transmitted and received images of Lena and show the quantitative measure of the state detection fidelity on the side in Fig.~\ref{figs6}.  Specifically, we can infer from Fig.~\ref{figs6}a and Fig.~\ref{figs6}b that neither medium turbulence nor the presence of  opaque particles significantly degrades the image in our protocol when the source coherence structure is employed for image encoding. The same conclusion applies to our protocol with the source coherence radius which is evidenced by Fig.~\ref{figs6}c.\\


